\documentclass[aps,prc,twocolumn,nofootinbib,amsmath,amssymb,showpacs,showkeys]{revtex4-1}
\usepackage{pifont}
\usepackage{graphicx}
\usepackage{color}
\usepackage{amsmath}
\usepackage{amssymb}
\usepackage[caption=false]{subfig}
\usepackage{multirow}

\newcommand{\bracket}[1]{{ \left< #1 \right> }}

\begin{document}

% To get line numbers using lineo package
% \linenumbers


\preprint{CMU/102-2012}

\title{Atmospheric Dependence of the Stopping Cosmic Ray Muon Rate
at Ground Level}



\author{G. Bernero}
\author{J. Olitsky}
\author{R. A. Schumacher}
\email[contact: ]{schumacher@cmu.edu}
\affiliation{Department of Physics, Carnegie Mellon University, Pittsburgh, PA 15213, USA}

%\thanks{}

%\collaboration{}%\noaffiliation


\date{\today}

%\begin{center}
%DRAFT - NOT FOR DISTRIBUTION - DRAFT
%\end{center}

\begin{abstract}

The rate of low energy ($\alt150$ MeV) cosmic ray muons was measured
at ground level as a function of several atmospheric parameters.
Stopped muons were detected in a plastic scintillator block and
correlations were determined using a linear regression model.  A
strong anti-correlation between fractional changes in the ground-level
pressure and stopping muon rate of $-3.0\pm0.5$ was found, and also a
$-4.1\pm0.5$ anti-correlation with the fractional change in
atmospheric height at 10 kPa pressure.  A weak positive correlation
with the 10 kPa temperature was also found, but it was shown not to be
statistically significant in our data set.  The same analysis was
applied to the total rate of all charged cosmic ray particles detected
with the same apparatus, and good agreement with previous work was
seen. The pressure and height correlation parameters for stopping
muons are larger than for the total rate of all charged particles by
factors of about 1.6 and 3.7, respectively.

\end{abstract}

%\pacs{PACS PACS PACS}
\pacs{96.50.S-, 96.50.sb, 94.20wq
     } % end of PACS codes
%\pacs{
%      {96.50.S-}{ Cosmics rays},
%      {96.50.sb}{ Energy spectra},
%      {94.20.wq}{ In ionosphere},
%     } % end of PACS codes

\keywords{cosmic rays, muons, atmosphere}

\maketitle

%\pagestyle{myheadings}
%\markboth{TEST 1}{TEST 2}

%*************************************************************************************
%                                                                                    *
%                                   Introduction                                     *
%                                                                                    *
%*************************************************************************************
\section{\label{intro}INTRODUCTION}

The rate of cosmic rays at ground level varies in time due to changes
in both the incident flux at the top of the atmosphere and changes in
atmospheric conditions. (For an overview see Ref.~\cite{Grieder}.)
Variations in the primary spectrum are thought to be due in part to
the indirect effect of the magnetospheric distortions caused by solar
activity, and these ``Forbush'' variations can occur on a timescale of
hours and days and be up to about 20\% in magnitude.  Separately, the
evolution of cosmic ray showers in the atmosphere is affected by gas
density variations, which are in turn related to measurable pressures
and temperatures at various altitudes, as has been discussed in the
literature for decades.  Incident protons between roughly 15 and 20 km
altitude initiate collisions with nuclei that produce primarily pions
and spallation nuclear fragments.  When this occurs at comparatively
higher altitude and at lower air density, the pions are more likely to
decay to muons rather than to reinteract to create lower energy pions
and nucleonic secondaries. Higher average energy decay muons, though
fewer in number, are less likely to suffer enough ionization energy
loss to stop and decay before reaching the ground.  On the other hand,
if the pionic interactions take place at comparatively lower height
and higher air density, pions are more likely to reinteract before
decaying, ultimately producing a larger number of lower energy muons
closer to the ground.  The interplay of these processes leads to the
well-known broad momentum spread of muons and other particles at
ground level~\cite{Beringer:1900zz}, but also to correlations of
particle rates and atmospheric conditions.

The ground-level air pressure and cosmic ray rate are anti-correlated,
since denser air causes more ionization energy loss to occur, and
therefore earlier stoppage of muons. Also, for a given temperature
lapse rate, denser air at ground level means a higher altitude at
which muons are formed, which in turn gives muons more chance to decay
before arriving at the surface.

The mean height of muon formation coincides with roughly the 10 kPa
(100 mb) pressure level. Again, an anti-correlation is expected with
changes in this height, since formation of muons at higher altitude
means that fewer will arrive at ground level.  Direct measurements of
the height of the atmosphere using balloon-type soundings are
available~\cite{IGRA,IGRAdata}.  We used the 10 kPa pressure-level
height to analyze our data.  We then studied the variability of the
results with respect to this choice, as will be shown.

The temperature of the atmosphere at the mean muon production altitude
may be supposed to affect the muon rate in the following way: if the
air is warmer it is less dense (at fixed pressure), so the pions have
greater probability of decaying to muons before they reinteract.  This
would favor creation of higher average energy muons that reach ground
level, and also lead to larger numbers that penetrate deep
underground.  A positive correlation with temperature is seen
underground, for example in Refs.~\cite{Barrett,Cini,Adamson:2012gt}.
But the expected effect at ground level is less definite, since lower
average energy muons created from pions in a denser atmosphere may be
more copious in number at the surface, formed together with fewer
higher energy muons.  Indeed the temperature correlation was too small
to measure in Ref.~\cite{Cotton} for vertically incident muons in a
sea-level telescope of counters.

Shower creation and propagation can be modeled~\cite{Dorman, Adamson:2012gt}
or simulated in detail, including the correlations among the various
parameters characterizing the atmosphere, but this was not the goal of
the present study.  Our goal was the experimental determination of the
correlation among several atmospheric parameters and the ground-level
stopping muon rate.  We also measured the parameters of the total
rate of charged particles to see whether they are measurably
different. The properties of the atmosphere that we examined in
relation to the particle rates were as follows.  First, we used the
ground-level air pressure, $P$, measured at the site of the detector.
Second, we used weather data giving the height of the atmosphere at
certain specific pressures, $H$, via public data~\cite{IGRA}.  Third,
from the same weather data set, we used the temperature of the
atmosphere, $T$, at certain specific pressures.  Finally, we looked
for correlations with the low-altitude humidity of the air, but these
were found to be nonexistent in this work.  Atmospheric pressure at
any altitude is proportional to air density, as is the inverse of air
temperature, and it is the density that ultimately regulates the
evolution of cosmic ray showers.  Our results were more satisfactory,
however, when looking for the separate correlations with the given
parameters.

In order to investigate the rate dependence of cosmic rays and
stopping muons, the following linear regression model was adopted.  As
discussed in the next section, a seven week span of time during which
the data did not exhibit drastic fluctuations in rate $\Phi$, was
defined by inspection of the data from a five month long run.  For
that span of time the average values of all variables in the data were
established (bracketed symbols).  The expression for relating the
variables was taken to be
\begin{equation} 
\frac{\Phi-\bracket{\Phi}}{\bracket{\Phi}} = \alpha
\left(\frac{P-\bracket{P}}{\bracket{P}}\right) + \beta 
\left(\frac{H-\bracket{H}}{\bracket{H}}\right) + \gamma
\left(\frac{T-\bracket{T}}{\bracket{T}}\right), 
\label{eq:regress}
\end{equation} 

\noindent
where $\alpha$, $\beta$, and $\gamma$ are unitless coefficients of
fractional change, and the other variables were defined above.  This
expression can be viewed as the linear part of the Taylor series that
should, to good approximation, capture whatever the effective
relationship among the given variables is.  A least-squares fit was
used to determine $\alpha$, $\beta$, and $\gamma$, and reduced
$\chi^2$ gave a measure of the goodness of the fits.  The parameters
are assumed to be uncorrelated for the purpose of comparing our
results to historical precedents for non-stopping muons.  Note that in
the literature there are various names and formulations for these
coefficients.  For example, $\alpha$ is sometimes called the
``negative partial barometer coefficient'' and reformulated in units
of \%/cm Hg.  If $\beta$ and $\gamma$ are not fitted, it is called the
``total barometer coefficient''.

Section~\ref{sec:approc} will discuss our equipment and procedures for
these measurements, followed by the results and comparison to other
work in Section~\ref{sec:results}.  Conclusions will be stated in
Section~\ref{sec:conclusions}.

\section{\label{sec:approc}APPARATUS AND PROCEDURES}

The detector for both stopping muons and for counting the total rate
of through-going cosmic rays is shown in Fig.1.  It consisted of a 30
cm cube of polystyrene scintillator, viewed from opposite sides by two
5'' diameter photomultipliers (PMT).  For vertically impinging
particles, this detector stopped muons of kinetic energy less than 100
MeV (177 MeV/c), while for oblique tracks the maximum energy was 150
MeV (233 MeV/c).  This defined what we mean by low-energy stopping
muons, of which we determined the rate variations.  There were no
separate trigger counters to select vertically-moving particles, so
this apparatus accepted the full-sky angular distribution of arriving
cosmic rays.  The equipment was located under a thin metal roof in a
building with a largely unobstructed view of the full sky.

%%%%%%%%%%%%%%%%%%%%%%%%%%Fig 1%%%%%%%%%%%%%%%%%%%%%%%%%%%%%%%%%%%%
%remove the ``*'' to put the figures in-line with text.  
%\begin{figure}
%\resizebox{0.50\textwidth}{!}{\includegraphics[angle=0.0]{apparatus.pdf}}
\begin{figure}
\resizebox{0.5\textwidth}{!}{\includegraphics[angle=-90.0]{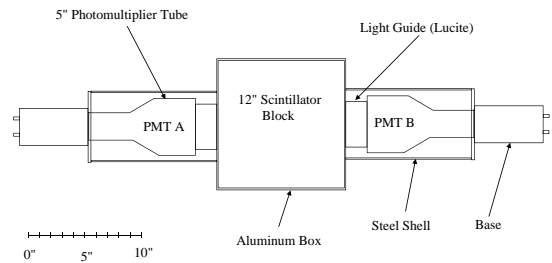}}
\vspace{-2.0cm}
\caption{Scintillator block and photomultiplier pair used to detect
  both stopping muons and to count passing cosmic rays.  }
\label{fig:apparatus}       % Give a unique label
\end{figure}
%\end{figure*} 
%%%%%%%%%%%%%%%%%%%%%%%%%%Fig 1%%%%%%%%%%%%%%%%%%%%%%%%%%%%%%%%%%%%

The signals from each photomultiplier were sent to discriminators
which formed logic pulses from all input signals larger than a pre-set
threshold of 120 mV.  The discriminator signals were sent to a
coincidence unit to create pulses only when signals from both tubes
were present within a 20 nsec time window, thus reducing noise and
allowing only one pulse per particle.  These pulses were used for
determining the muon lifetimes with use of a time to digital converter
(TDC). The data were accumulated event-by-event using a National
Instruments data acquisition (DAQ) unit (Model NI USB-6221) and a
LabVIEW-based control system.  A stopping muon candidate was defined
by a pulse pair with a TDC ``Start'' pulse followed by any ``Stop''
pulse within 20 $\mu$sec.  Pulse pairs with a longer time interval
were plentiful, caused by uncorrelated cosmic ray particles, and these
were counted but not individually recorded.  The PMT singles rates
were $\sim 300$ Hz, the coincidence total count rate for the
experiment averaged 24/sec, while the rate of stopping candidates was
about 6/min.  For each candidate stopping muon event, the following
parameters were recorded: a time-stamp, the decay time in $\mu$sec
with 12.5 nsec (80 MHz) time resolution, three redundant pressure
readings, and the total of cosmic ray event pairs detected up to that
moment.

The ground-level pressure was measured simultaneously with the cosmic
ray data using the same LabVIEW-based DAQ system.  We used three
single-chip barometers (Motorola MPX4115A) to record the air pressure
whenever a stopped-muon event occurred.  The average of the three
readings was used as the final value for the pressure.  We used METAR
weather data from the Allegheny County Airport (KAGC) located 16 km
away, in order to check the accuracy of the lab pressure data.

The atmospheric data at high altitude was obtained using publicly
available Integrated Global Radiosonde Archive (IGRA)
data~\cite{IGRAdata}.  A parser of the extensive ASCII data set was
written to extract the needed pressure, altitude, temperature, and
time information.  The data from the in-lab detectors and the
atmospheric data were merged together chronologically in the off-line
analysis.  The IGRA data was available in 12 hour intervals. Initially
we binned the muon data in one-hour intervals to get the average
pressure and the number of both stopping and total events.  These data
were then rebinned, as needed, to match the available IGRA atmospheric
data.  Temperature and atmospheric height data were available in this
data set at only a small set of pressures.  One of them was 10 kPa,
which occurs at about 16.6 km altitude, where primary cosmic rays have
started to shower.  This was the pressure level selected by us to fit
our model.

The experiment ran for 21 weeks from July 5, 2011 to December 1, 2011,
with some gaps for equipment repair.  A total of $1.31\times10^6$
stopping muons were detected in this time, for an average rate of
close to 350 per hour.  At the low muon momenta considered in this
measurement, there is a slight ($\sim10$\%) excess of $\mu^+$ over
$\mu^-$ particles~\cite{Beringer:1900zz}.  Our apparatus did not
distinguish between the two.  Stopped $\mu^-$ events are known to
undergo atomic capture on carbon atoms in the scintillator, leading to
a reduced lifetime due to the weak $\mu^- + p \rightarrow \nu_\mu + n$
process.  The value is $2.028\pm0.002$ $\mu$sec ~\cite{Measday}.  The
$\mu^+$ events decay with their free-space lifetime of
$2.1969811\pm0.0000022$ $\mu$sec~\cite{Beringer:1900zz}.  The measured
muon lifetime from the effective exponential decay distribution was
$2.117\pm0.003$ $\mu$sec, between the two other values, as expected.
This verified that our data sample consisted of muons that stopped and
decayed. Within the 20~$\mu$sec time window for selecting stopped
muons, the fraction of uncorrelated non-stopping background track
pairs was 11.5\%.  This background was subtracted in the following
stopped muon analysis.

\section{\label{sec:results}RESULTS}

\begin{figure*}
\vspace{-3.0cm}
\resizebox{1.0\textwidth}{!}{\includegraphics[angle=0.0]{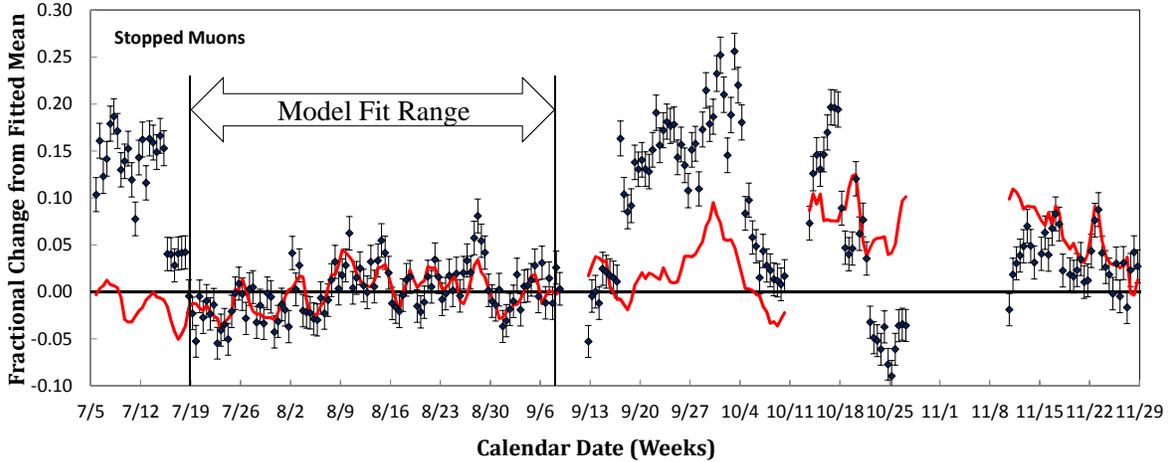}}
\vspace{-4.0cm}
\caption{ (Color online) 
Twenty-one weeks of stopping muon rate in 12 hour intervals expressed
as a fractional change from the mean in the Model Fit Range.  The
error bars are statistical.  The solid red curve is the result of a
fit to the data in the Model Fit Range.  }
\label{fig:stoppeddata_full}       % Give a unique label
%\end{figure}
\end{figure*}

Figure~\ref{fig:stoppeddata_full} shows the resultant fractional rate
variation of stopping muon candidates over the entire data period,
together with the model fit discussed below.  The rate is presented as
the fractional deviation from the mean of the fitted period between July
19 to September 8, 2011.  The average variation
during this time is zero, by construction. The error bars are purely
statistical.  One sees day-to-day fluctuations of about $\pm2.5$\%, as well
as larger excursions outside the fitted region of up to 20\%.  

%%%%%%%%%%%%%%%%%%%%%%%%%% Fig 2 %%%%%%%%%%%%%%%%%%%%%%%%%%%%%%%%%%%%
%remove the ``*'' to put the figures in-line with text.  
%\begin{figure}
%%%%%%%%%%%%%%%%%%%%%%%%%% Fig 2 %%%%%%%%%%%%%%%%%%%%%%%%%%%%%%%%%%%%

\begin{figure*}
\vspace{-3.0cm}
\resizebox{1.0\textwidth}{!}{\includegraphics[angle=0.0]{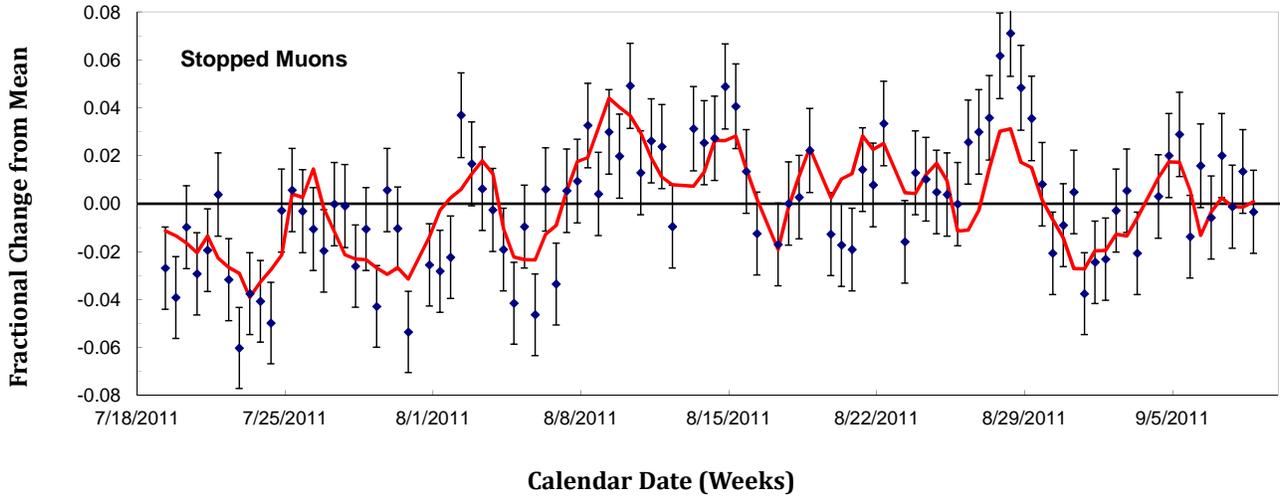}}
\vspace{-3.0cm}
\caption{ (Color online) 
Detail of data from the Model Fit Range in
Fig.~\ref{fig:stoppeddata_full}, showing the rate dependence of the
stopping muons together with the best-fit result to the model
discussed in the text.  The error bars are statistical.  }
\label{fig:stoppeddata_fit}       % Give a unique label
%\end{figure}
\end{figure*}

In seeking correlations with atmospheric data, it was found that the
large jumps visible in the second week and again after the eleventh
week were impossible to match.  We ascribed these large jumps to
changes in ``space weather'' of the Forbush variety that were outside
of our scope to track and model.  An interval of data that was
comparatively devoid of drastic fluctuations in rate was used to fit
to the model.  This was the seven week period marked in the figure as
``Model Fit Range''.  The data and the model fit
(Eq.~\ref{eq:regress}) from this restricted range of data are shown in
Fig.~\ref{fig:stoppeddata_fit}.  The solid red curve in both figures
is the result of the regression model fit.  Each of the three
parameters in the model is a unitless scale factor relating a
fractional change in atmospheric condition to a fractional change in
particle rate, as seen in Eq.~\ref{eq:regress}.  A least-squares fit
optimized parameters $\alpha$ for the ground-level pressure, $\beta$
for the 10 kPa atmospheric height, and $\gamma$ for the 10 kPa
temperature.  The resultant values of those parameters are given in
Table~\ref{resulttable1}.  The uncertainties on the parameters come
from the diagonal elements of the error matrix associated with the
least-squares fit.  Our data fits the regression model ansatz with
good agreement, as demonstrated by the reduced chi-squared
$\chi^2_\nu$ value of 1.07, allowing an rms spread of $\sqrt{2/\nu} =
0.15$.  The Pearson's correlation coefficient is 0.73, which is
significant in view of the large statistical uncertainties on the data
points.

%%%%%%%%%%%%%%%%%%%%%%%%%% Fig 3 %%%%%%%%%%%%%%%%%%%%%%%%%%%%%%%%%%%%
%remove the ``*'' to put the figures in-line with text.  
%\begin{figure}
%%%%%%%%%%%%%%%%%%%%%%%%%% Fig 3 %%%%%%%%%%%%%%%%%%%%%%%%%%%%%%%%%%%%

%%%%%%%%%%%%%%%%%%%%%%%%%% Table 1 %%%%%%%%%%%%%%%%%%%%%%%%%%%%%%%%%%
\begin{table}
\caption{Model parameters for the correlation of particle rates with
  ground-level pressure ($\alpha$), the 10 kPa height of the atmosphere
  ($\beta$), and the 10 kPa temperature ($\gamma$).  The third column is
  for low energy muons stopping at ground level, while the fourth
  column is for the total detected cosmic particle rate.}
\begin{center}
\begin{tabular}{cccc}
\hline
\hline
Parameter   & Eq.~\ref{eq:regress} &Stopped      & Total Particles    \\ 
            &                      &Muons        &                    \\ 
\hline 
Pressure    & $\alpha$ & $ -3.2 \pm 0.5 $        & $-1.94  \pm 0.10$  \\ 
Altitude    & $\beta $ & $ -2.7 \pm 0.9 $        & $-0.8   \pm 0.2$   \\ 
Temperature & $\gamma$ & $+0.35 \pm 0.17$        & $+0.08  \pm 0.04$  \\ 
\hline 
            &          &$\chi^2_\nu$ = 1.07      & $\chi^2_\nu $= 1.09\\ 
\hline
\hline
\end{tabular}
\end{center}
\label{resulttable1}
\end{table}
%%%%%%%%%%%%%%%%%%%%%%%%%% Table 1 %%%%%%%%%%%%%%%%%%%%%%%%%%%%%%%%%%

%%%%%%%%%%%%%%%%%%%%%%%%%% Table 2 %%%%%%%%%%%%%%%%%%%%%%%%%%%%%%%%%%
\begin{table}
\caption{Model parameters for the correlation of particle rates as in
  Table~\ref{resulttable1}, but with only two free parameters. 
  The third column is for muons stopping at
  ground level, while the fourth column is for the total detected
  cosmic particle rate.}
\begin{center}
\begin{tabular}{cccc}
\hline
\hline
Parameter   & Eq.~\ref{eq:regress} &Stopped      & Total Particles    \\ 
            &                      &Muons        &                    \\ 
\hline 
Pressure    & $\alpha$ & $ -3.0 \pm 0.5 $        & $-1.9   \pm 0.1$   \\ 
Altitude    & $\beta $ & $ -4.1 \pm 0.5 $        & $-1.1   \pm 0.1$   \\ 
\hline 
            &          &$\chi^2_\nu$ = 1.11      & $\chi^2_\nu $= 1.13\\ 
\hline
\hline
\end{tabular}
\end{center}
\label{resulttable2}
\end{table}
%%%%%%%%%%%%%%%%%%%%%%%%%% Table 2 %%%%%%%%%%%%%%%%%%%%%%%%%%%%%%%%%%

%%%%%%%%%%%%%%%%%%%%%%%%%% Table 3 %%%%%%%%%%%%%%%%%%%%%%%%%%%%%%%%%%
\begin{table}
\caption{Model parameters for the correlation of particle rates as in
  Table~\ref{resulttable1}, but with only one free parameter. 
  The third column is for muons stopping at
  ground level, while the fourth column is for the total detected
  cosmic particle rate.}
\begin{center}
\begin{tabular}{cccc}
\hline
\hline
Parameter   & Eq.~\ref{eq:regress} &Stopped      & Total Particles    \\ 
            &                      &Muons        &                    \\ 
\hline 
Pressure    & $\alpha$ & $ -3.3 \pm 0.5 $        & $-2.0   \pm 0.1$   \\ 
\hline 
            &          &$\chi^2_\nu$ = 1.74      & $\chi^2_\nu $= 2.22\\ 
\hline
\hline
\end{tabular}
\end{center}
\label{resulttable3}
\end{table}
%%%%%%%%%%%%%%%%%%%%%%%%%% Table 3 %%%%%%%%%%%%%%%%%%%%%%%%%%%%%%%%%%

The visual appearance of the fit within the fitted range is good. We
note that there is an anti-correlation of surface pressure with
stopping muons rate, since $\alpha$ is large and negative.  There is
an anti-correlation of similar size with atmospheric height, as given
by $\beta$.  Also, there is a weaker positive correlation with
temperature at the 10 kPa level, as given by $\gamma$ with a fairly
large uncertainty.

We tested the significance of the fit by repeating it with fewer
parameters.  The result with the temperature factor $\gamma$ excluded
is shown in Table~\ref{resulttable2}.  The quality of the fit, as
estimated by the reduced $\chi^2$, is almost the same, so we can
conclude that the temperature of the atmosphere at the altitude of 10
kPa pressure is a statistically insignificant variable.  Parameter
$\alpha$ hardly changes, but $\beta$ gets larger.  There was scarcely
any visual difference between the curves produced by these two fits.
These resultant values are our best result for this measurement.

To further test the model we restricted the parameters again to use
only $\alpha$ for the ground-level pressure, excluding the height
parameter $\beta$.  This is shown in Table~\ref{resulttable3}.  The
quality of the fit is now significantly worse, as seen in the reduced
$\chi^2$, and the visual appearance of the fit was poor.  Thus, ground
level pressure alone is not enough to accurately track the rate of
stopping muons.

%%%%%%%%%%%%%%%%%%%%%%%%%% Fig 4 %%%%%%%%%%%%%%%%%%%%%%%%%%%%%%%%%%%%
%remove the ``*'' to put the figures in-line with text.  
%\begin{figure}
%\resizebox{0.50\textwidth}{!}{\includegraphics[angle=0.0]{apparatus.pdf}}
\begin{figure*}[htpb]
\vspace{-3.0cm}
\resizebox{1.0\textwidth}{!}{\includegraphics[angle=0.0]{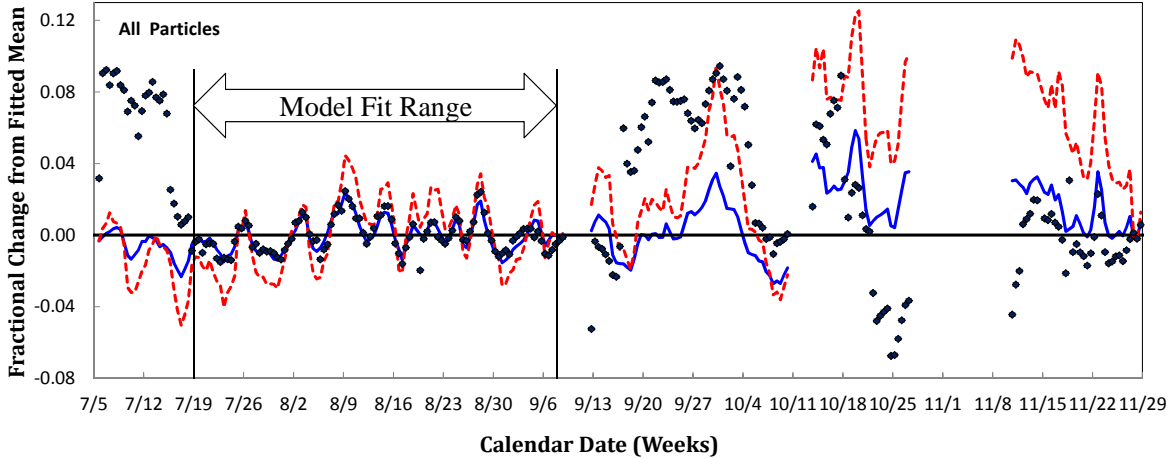}}
\vspace{-4.0cm}
\caption{ (Color online) 
For the same time period as in Fig.~\ref{fig:stoppeddata_full}, the total
cosmic particle rate expressed as a fractional change from the mean in
the Model Fit Range.  The error bars are purely statistical and
generally smaller than the symbols.  The solid blue curve is the
result of a fit to the data within the Model Fit Range.  The dashed
red curve is the fit for the stopped muons from
Fig.~\ref{fig:stoppeddata_fit}. }
\label{fig:totaldata_full}       % Give a unique label
%\end{figure}
\end{figure*}
%%%%%%%%%%%%%%%%%%%%%%%%%% Fig 4 %%%%%%%%%%%%%%%%%%%%%%%%%%%%%%%%%%%%

We used the same method to fit our data for the total event rate.  This
total rate at ground level in this detector consists mostly of higher
momentum muons (that do not stop), but with minor contributions from
electrons, protons, and neutrons.  The total rate includes the very
small fraction ($\sim 0.4\%$) of stopped muon rate.  The resultant fit
is shown in Fig.~\ref{fig:totaldata_full}, where one sees immediately
the effect of the much higher statistics when counting all particles
rather than just the lowest energy muons.  The fractional change in
total particle rate is now seen to vary quite smoothly on the
timescale of days and weeks, albeit with larger variations of order
10\%.  There are two model curves superimposed.  The dashed red curve
uses parameters from Table~\ref{resulttable1} obtained for the fit to
stopped muons as per Fig.~\ref{fig:stoppeddata_fit}.  The solid blue
curve shows the new fit to the total particle rate with parameters
also shown in Table~\ref{resulttable1}.  Within the Model Fit Range,
the solid blue fit line is a close match to the data.  Outside this
range one sees again the large fluctuations we ascribe to space
weather conditions that we could not track or reproduce.

%%%%%%%%%%%%%%%%%%%%%%%%%% Fig 5 %%%%%%%%%%%%%%%%%%%%%%%%%%%%%%%%%%%%
%remove the ``*'' to put the figures in-line with text.  
%\begin{figure}
%\resizebox{0.50\textwidth}{!}{\includegraphics[angle=0.0]{apparatus.pdf}}
\begin{figure*}
\vspace{-3.0cm}
\resizebox{1.0\textwidth}{!}{\includegraphics[angle=0.0]{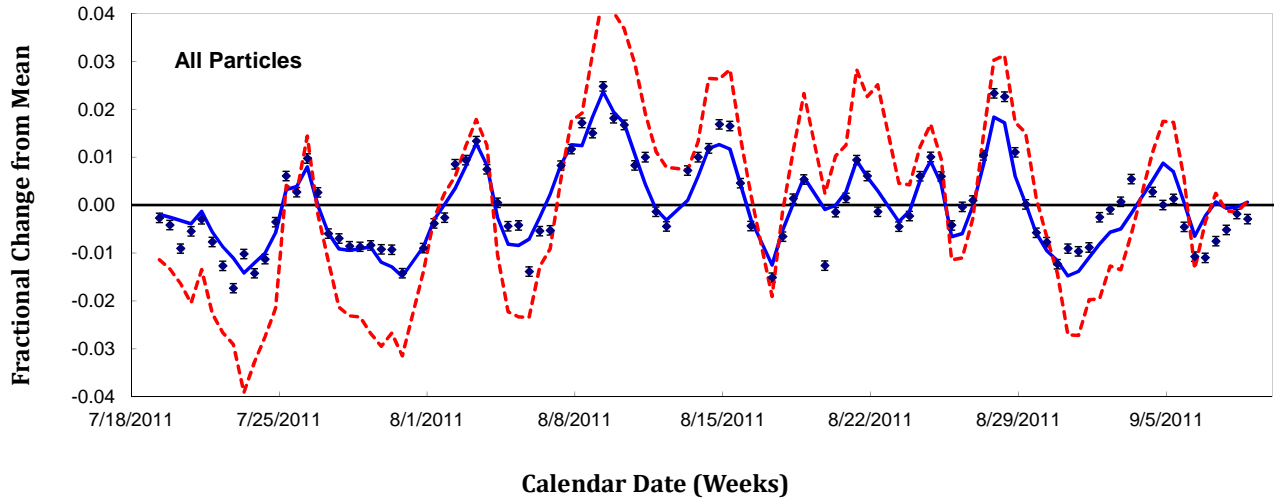}}
\vspace{-4.0cm}
\caption{ (Color online) 
Detail of data from the Model Fit Range in
Fig.~\ref{fig:totaldata_full}, showing the total cosmic particle rate
expressed as a fractional change from the mean.  The error bars are
purely statistical.  The curves are as in
Fig.~\ref{fig:totaldata_full}.  }
\label{fig:totaldata_fit}       % Give a unique label
%\end{figure}
\end{figure*}
%%%%%%%%%%%%%%%%%%%%%%%%%% Fig 5 %%%%%%%%%%%%%%%%%%%%%%%%%%%%%%%%%%%%

In fact, the fit to the regression model within the fit range led to a
large value of $\chi^2_\nu$ of about 15.  The statistics of this data
set was so high that small fluctuations of unknown origin, presumed to
be ``space weather'', were evident.  In order to compute legitimate
uncertainties on the fit parameters in the least-squares algorithm, we
increased the point-to-point uncertainties from 0.0010 to 0.0035 to
account for these fluctuations.  This procedure did not affect the
parameter values but allowed for the estimation of reasonable
uncertainties for the correlation parameters.  The fit result in the
Model Fit Range is shown in Fig.~\ref{fig:totaldata_fit}, showing data
points with their unmodified statistical uncertainties.  The dashed
blue curve fits the data very well.  In this case the Pearson's
correlation coefficient is 0.92, also indicating a very good fit.

It can be seen that the amplitude of the fractional changes of the
stopped muon rate (dashed red curve) is larger than the amplitude of
the total particle rate changes (solid blue curve). However, they
track each other very well over time.  The fit parameters for the
total particle rate are included in Tables~\ref{resulttable1},
~\ref{resulttable2} and ~\ref{resulttable3}.  We see that $\alpha$ is
about 50\% larger for the stopped muons, while $\beta$ is about four
times as large.  Taking into account the uncertainties on the values,
$\alpha$ is larger for stopping muons at the ``1.8 sigma'' level, and
$\beta$ at the ``5 sigma'' level in Table~\ref{resulttable2}.  From
the figure, there can hardly be any doubt that the atmospheric
dependencies indeed differ between the total particle rate and the
stopping muon rate.  Thus we conclude that the lowest energy cosmic
ray muons, the ones that stop in this detector, are several times more
sensitive to ground-level pressure and atmospheric height variations
than the total particle rate.

It will be noted that our results do not include the ground
level humidity measurements.  Using local METAR data we found that the
drastic variations in humidity from day to day and week to week were
not correlated in any detectable way to the stopping muon rate.  Thus,
we simply report that there was no measurable correlation.

\begin{figure}
%\begin{figure*}
\resizebox{0.5\textwidth}{!}{\includegraphics[angle=0.0]{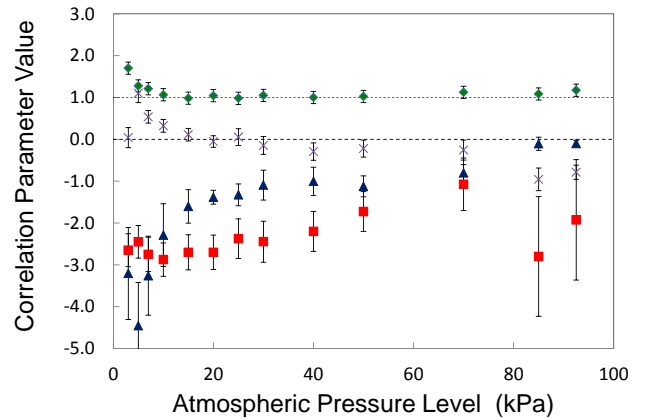}}
\vspace{-1.0cm}
\caption{ (Color online) 
Evolution of the correlation parameters for stopped muons as a
function of the reference pressure level (in kPa) used for atmospheric
height and temperature.  Solid red squares: ground-level pressure
parameter $\alpha$, solid blue triangles: atmospheric height $\beta$,
crosses: atmospheric temperature $\gamma$.  The solid green diamonds
are the reduced $\chi^2$ values for the fits at each pressure level.
}
\label{fig:heighttrend}       % Give a unique label
\end{figure}

Returning to results for the stopped muons, we assumed initially that
the height and the temperature of the atmosphere at 10 kPa pressure
was the optimal single level at which to determine the correlations.
This was based on the knowledge~\cite{Grieder} that this pressure
level in the atmosphere is dominant in the formation of cosmic ray
showers.  However, one clearly expects the full development of muon
showers to depend on the density of the atmosphere throughout an
extended region, for which a full theoretical and numerical treatment
is required~\cite{Dorman}.  Within the limited scope of our work we
selected other reference altitudes from the IGRA data at which to
compute the correlations for stopped muons to test the 10 kPa
assumption.  The results are shown in Fig.~\ref{fig:heighttrend}.  As
can be seen, the $\chi^2_\nu$ values remain statistically close to
unity except for the high and low extremes of altitude. This suggests
that the parametrization we used works about equally well from $\sim
5$ to $\sim 70$ kPa, albeit with differing values for $\alpha$,
$\beta$, and $\gamma$, wherein no single altitude seems more sensitive
than the others. We believe that the fit worsens at the high altitude
limit, below 10 kPa, because the height and temperature there are
increasingly above where the atmosphere has affected the primary
cosmic particles. We believe that the increased uncertainties at the
lowest altitudes are due to a breakdown of our assumption that the
pressure measured in the laboratory is independent of the parameters
measured by the IGRA radiosondes. The model fails at low altitudes
where the quantities are closely correlated.  We note in the figure
that the pressure parameter $\alpha$ tends to be smaller at lower
altitudes.  This is consistent with the trend reported in an early
study for the total particle rate discussed in Ref.~\cite{Janossy}.

%%%%%%%%%%%%%%%%%%%%%%%%%% Fig 6 %%%%%%%%%%%%%%%%%%%%%%%%%%%%%%%%%%%%
%remove the ``*'' to put the figures in-line with text.  
%\end{figure*}
%%%%%%%%%%%%%%%%%%%%%%%%%% Fig 6 %%%%%%%%%%%%%%%%%%%%%%%%%%%%%%%%%%%%

%%%%%%%%%%%%%%%%%%%%%%%%%% Table 4 %%%%%%%%%%%%%%%%%%%%%%%%%%%%%%%%%%
\begin{table*}
\caption{Comparison of the present results to selected previous measurements.  
All historic values were converted to consistent unitless coefficients.
 }
\begin{center}
\begin{tabular}{ccccccccc}
\hline
\hline
Parameter   & Eq.~\ref{eq:regress} &Bernero           & Trumpy           &Chasson            & Shamos           & Cotton           & Grieder          & Fenton           \\ 
            &                      &(This Paper)      &Ref.\cite{Trumpy} &Ref.\cite{Chasson} &Ref.\cite{Shamos} &Ref.\cite{Cotton}  &Ref.\cite{Grieder}&Ref.\cite{Fenton} \\ 
\hline 
Stopped Data & & & & & & \\
\hline 
Pressure    & $\alpha$             & $ -3.2 \pm 0.5 $ & $-3.5 \pm 0.3 $  & $-3.6 \pm 0.5$    & $-7.1$           &                   &                  &                  \\    
Altitude    & $\beta $             & $ -2.7 \pm 0.9 $ & $0.0  \pm 0.3 $  & $-$               & $-$              &                   &                  &                  \\ 
Temperature & $\gamma $            & $ +0.35\pm 0.17$ & $+0.17\pm 0.17$  & $-$               & $-$              &                   &                  &                  \\ 
\hline 
Total Rate & & & & & & \\
\hline 
Pressure    & $\alpha$             & $ -1.94\pm 0.10$ & $-1.95\pm 0.07 $ & $-1.2 \pm 0.2$    & $-3.0$           & $-1.03 \pm 0.11$  & $-1.63    $      & $-0.45 \pm 0.03 $\\ 
Altitude    & $\beta $             & $ -0.8 \pm 0.2 $ & $-0.53\pm 0.08 $ & $-0.54 \pm 0.08$  & $-$              & $-0.52 \pm 0.14$  & $-0.83    $      & $-0.08 \pm 0.04 $\\ 
Temperature & $\gamma $            & $ +0.08\pm 0.04$ & $+0.10\pm 0.04 $ & $+0.14 \pm 0.04$  & $-$              & $+0.048\pm0.056$  & $\sim 0.2 $      & $+0.04 \pm 0.02 $\\ 
\hline
\hline
\end{tabular}
\end{center}
\label{resulttable4}
\end{table*}
%%%%%%%%%%%%%%%%%%%%%%%%%% Table 4 %%%%%%%%%%%%%%%%%%%%%%%%%%%%%%%%%%

In comparing the results of this measurement to previous work, we have
always converted the older results to our preferred unitless
parameterizations for $\alpha, \beta$ and $\gamma$ as in
Eq.~\ref{eq:regress}.  We used 16.6 km as the mean 10 kPa height of the
atmosphere and 207 K as the mean temperature at this height. One of
the previous experiments comparable to this one was performed by
Trumpy and Trefall~\cite{Trumpy}.  At ground level, they used stacks
of counters separated by 10 or more centimeters of lead to record the
total cosmic ray rates, and correlated these rates to the same
atmospheric parameters as we do in these measurements.
Table~\ref{resulttable4} includes their data for the total rate.  One
sees that their results agree with ours within errors for the total
particle rate.  They then used a model to subtract the ``hard''
component, defined as tracks passing through at least 10 cm of lead,
to arrive at a ``soft'' component.  This soft component is not the
same as our stopped muon measurement, but it is related.  Additional
``soft'' contributions to their result could arise from electrons from
in-flight muon decay, electrons from knock-on processes, and from very
slow hadrons of all sorts.  Nevertheless, we can compare their
``soft'' component with our results, and this is also shown in
Table~\ref{resulttable4}.  The values for the pressure correlation
$\alpha$ are in agreement, but they were unable to obtain meaningful
results for $\beta$ and $\gamma$.

Another of the scarce measurements of the ``soft'' cosmic ray
correlations was by Chasson~\cite{Chasson}.  Using stacks of Geiger
counter telescopes with and without lead and iron absorbers between
detectors, he factored $\alpha$ into ``hard'' and ``soft''
components. His result is included in Table~\ref{resulttable4}.  For
the total particle rate his results are in good agreement with ours,
within uncertainties.  For the stopped muon data he did not measure
all three parameters, but gave only a value for the total barometer
version of coefficient $\alpha$.  It is in agreement with ours.  He
did not measure stopping muons directly, as we have.

We also mention the relevant results of Shamos and
Liboff~\cite{Shamos}. On the ocean surface they measured using a stack
of ionization counters, so that, with some model assumptions, they
factored $\alpha$ into ``hard'' and ``soft'' components. Their result
is included in Table~\ref{resulttable4}.  Unfortunately, they did not
estimate uncertainties and they measured only the ``total''
coefficient, not the ``partial'' coefficients as we have.  Thus, the
poor agreement is difficult to evaluate.  Again, they did not measure
stopping muons directly, as we have.

Results from Cotton and Curtis~\cite{Cotton} show poor to fair
agreement with our results for the total particle rate.  Their
measurements were made at ground level using a stack of counters to
select only the vertical component of the cosmic rate.  Note that
their temperature correlation is compatible with zero.  Taking our
experiment and these earlier experiments together, all give barely
significant positive values for $\gamma$; thus, there may be a hint
for a small positive temperature correlation for the total particle
rate at ground level.

The text by Grieder~\cite{Grieder} offers a set of total rate
correlation parameters of unstated provenance and without uncertainty
estimates which are in the same range as the other results in the
table, including ours.

For comparison of ground level and underground measurements we
consider results from Fenton {\it et al.}~\cite{Fenton} taken at 42
meters water equivalent depth, shown in the last column of
Table~\ref{resulttable4}.  For those muons, estimated to have at least
15 GeV energy, the correlations with mean production height and
surface pressure are smaller than our results by roughly an order of
magnitude, and their small temperature coefficient is consistent
within errors.  At greater depths it is has been
shown~\cite{Adamson:2012gt} that the atmospheric effective temperature
is the main correlator with high energy muon rate.  Our small value
for parameter $\gamma$ at the surface is thus consistent with this
picture.

\section{\label{sec:conclusions}CONCLUSIONS}

We have measured the rate of stopping muons at the surface of the
earth and compared it with the rate of all charged particles.  As was
shown in Tables~\ref{resulttable1}, ~\ref{resulttable2} and
\ref{resulttable3}, the atmospheric coefficients for the low energy
stopping muons are significantly larger than for all particles.  The
linear regression model we have used represents both data sets well,
in the sense that the fits to the data capture the behavior over a
several-week period of low overall cosmic ray fluctuations, with good
fidelity.  The results are consistent with the picture that low energy
muons, the least penetrating portion of the cosmic ray spectrum, are
most sensitive to variations in atmospheric pressure, $P$, and the
upper-level density related to $H$.  We found no significant
correlation with temperature at the mean production height, estimated
using $T$.  Our results were compared to previous measurements and
found to be in fair to good agreement for the total particle rate.  We
were unable to find any previous measurements of the correlations for
the stopping muon rate.

It is also clear, as seen in Figures~\ref{fig:stoppeddata_full} and
\ref{fig:totaldata_full}, that there are large and sudden departures
from this model, which we ascribe to changes in the rate of incoming
primary cosmic rays, albeit without independent evidence that this was
the case.  In the same figures it is seen that the model trends were 
sometimes qualitatively consistent with the rate changes in time
periods not included in the fit region.  However, certain time
intervals exhibited unexplained step-like transitions in particle
rates not accounted for in our phenomenological atmospheric weather
data fit.  Overall, we believe our most significant result is the set
of correlation parameters for the rate of low energy muons that are
stopped at ground level.  We have shown that these correlations are
significantly larger than those for the total rate of all cosmic ray
particles.

\begin{acknowledgments}
We thank Professor Barry Luokkala and Mr. Albert Brunk for support in
obtaining some of the equipment needed for this study.
\end{acknowledgments}
\vfill

%-------------------------------------------------------

\vfill

\end{document}